Orders-of-magnitude speedup in atmospheric chemistry modeling through neural

network-based emulation

Makoto M. Kelp[a], Christopher W. Tessum[a]*, and Julian D. Marshall[a]

[a]University of Washington Department of Civil and Environmental Engineering, Box

352700, Seattle, WA 98195-2700, USA

* Corresponding author. Tel: (206) 543-2390. E-mail address: ctessum@uw.edu

**ABSTRACT**

Chemical transport models (CTMs), which simulate air pollution transport, transformation, and removal, are computationally expensive, largely because of the computational intensity of the chemical mechanisms: systems of coupled differential equations representing atmospheric chemistry. Here we investigate the potential for machine learning to reproduce the behavior of a chemical mechanism, yet with reduced computational expense. We create a 17-layer residual multi-target regression neural network to emulate the Carbon Bond Mechanism Z (CBM-Z) gas-phase chemical mechanism. We train the network to match CBM-Z predictions of changes in concentrations of 77 chemical species after one hour, given a range of chemical and meteorological input conditions, which it is able to do with root-mean-square error (RMSE) of $\leq$ 1.97 ppb (median RMSE = 0.02 ppb), while achieving a 250× computational speedup. An additional 17× speedup (total 4250× speedup) is achieved by running the neural network on a graphics-processing unit (GPU). The neural network is able to reproduce the emergent behavior of the chemical system over diurnal cycles using Euler integration, but additional work is needed to constrain the propagation of errors as simulation time progresses.

**1. INTRODUCTION**:

Limitations in computational speed and availability often limit the rate of scientific progress. In the physical sciences, computationally intensive numerical solutions to differential equations are a common cause of bottlenecks. The prevalence of legacy software unable to take full advantage of modern computer hardware can compound this limitation.

Air pollution modeling is one field that is often constrained by computational bottlenecks. Models exist with detailed representations of atmospheric chemistry and physics,[1-4] but they are computationally intensive enough to preclude robust quantification of model uncertainty.[5] Demand for air quality model results with reduced computational expense is evident from the wide variety of alternative air quality modeling techniques, such as simple heuristics,[6-8] statistical emulators of specific model outputs,[9] simplified versions of the full air quality models,[10, 11] and extensive mathematical reformulations of the models for specific types of analyses.[12-14] However, none of these alternatives can fully replace comprehensive models.

The most computationally expensive operation in an air pollution model is typically the chemical mechanism,[15, 16] which models reactions among pollutants by numerically integrating systems of ordinary differential equations (coupled with equations representing instantaneous phenomena) through time. Many of the reactions occur at short time scales, rendering them computationally intensive to explicitly model, especially because the typical simulation time considered is on the order of several weeks to a year. The chemical mechanism—and especially its aerosol chemistry subcomponent—is also typically among the most complex parts of the model software, making it difficult to optimize for use with modern hardware such as graphics processing units (GPUs).

We investigate an alternative to the explicit mechanistic modeling currently used in chemical mechanisms and other parts of physical models: emulation. We aim to test whether training an emulator to reproduce the behavior of a model component—such as a chemical mechanism—can both reduce the computational intensity of the operation and allow it to run on a GPU, decreasing the time required to perform the operation by orders of magnitude.[17, 18] Specifically, we present and evaluate a neural-network based emulator of a contemporary chemical mechanism. This research is part of the CACES (Center for Air, Climate, and Energy Solutions) EPA-ACE Center.

Previous work has explored the use of neural networks to solve differential equations,[19-21] but to our knowledge machine learning and neural networks have not previously been used to emulate a mechanism of the complexity presented in this work. In the physical sciences, machine learning and neural networks have been used to predict emergent model results or future observations[22-42] but not as a replacement for an internal model component (chemistry) as attempted here.

## 2. MATERIALS AND METHODS:

### 2.1 Chemical mechanism

We emulate the Carbon Bond Mechanism Z (CBM-Z;[43, 44]), which simulates tropospheric gas-phase chemistry. CBM-Z is included as a component in the WRF-Chem air quality model[45] but for our experiments we use a standalone version of the model; this version simulates the atmosphere as a single homogenous box (i.e., a "one-compartment model"). We configure the CBM-Z box model as shown in Table S1, leaving as free parameters initial conditions of 77 pollutant species and the cosine of the solar zenith angle. We use values for temperature, pressure, and relative humidity that are constant across all simulations.

### 2.2 Emulator Architecture

The CBM-Z model is a non-linear function with multiple continuous input variables and multiple continuous output variables. This type of function can be empirically modeled using multi-target regression, where "regression" refers to continuous rather than categorical outputs and "multi-target" refers to more than one output. Neural networks[46, 47] are one class of machine learning models that can perform multi-target regression.

Figure 1 displays the model architecture of the neural network employed in this work. Inputs entering the network are normalized using the mean and standard deviation of the training dataset, then fed through a single fully-connected layer followed by four residual blocks[48] with two fully-connected layers to each block, where the number of nodes in each block is equal to the number of CBM-Z input variables (i.e., 78). Each fully-connected layer is followed by batch normalization[49] and a rectified linear unit (ReLU)[50]. The final residual block is connected to an output layer with number of nodes equal to the number of CBM-Z chemical species (i.e., 77). We do not perform any regularization in addition to batch normalization, instead relying on a large training dataset to avoid overfitting.

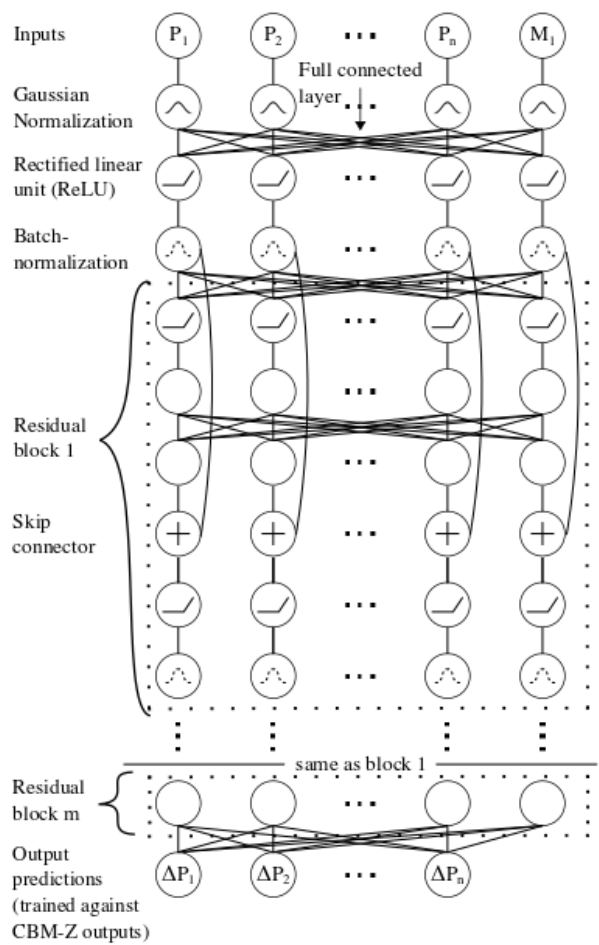

Fig. 1. Neural network architecture with *m* residual blocks for emulating a chemical mechanism with *n* pollutants ($P_n$) and one meteorological input ($M_1$). The network ouputs predicted changes in pollutant concentrations ($\Delta P_n$) over a specified time period.

## 2.3 Neural Network Training

We run the CBM-Z model to create a training dataset of 100 million examples, with each example including input pollutant concentrations, meteorology conditions, and output changes in pollution concentrations after one hour of CBM-Z simulation time, which corresponds to 12 5-minute time steps.

Training examples are created by 1) using Latin hypercube sampling to generate pseudorandom values for each variable within the typical atmospheric ranges described in Table S2; 2) initializing CBM-Z with these random values and running it forward for 3 hours of simulation time (i.e. a 'spin-up') to allow the system to adjust to a state closer to what might occur in ambient air; and 3) running the model forward for 24 hours of simulation time to simulate a diurnal cycle of air pollution, exogenously adjusting the solar zenith angle as the model runs, to represent the changing location of the sun. For each diurnal cycle, we save 24 examples, with each example representing the initial conditions at the beginning of an hour and the concentration changes over the course of the hour. We also create a separate evaluation dataset of 8 million examples using the same methodology. Training and evaluation examples are fed randomly into the neural network: it is not given information regarding any temporal relationships among examples.

We train the neural network by minimizing average mean-squared-error (MSE) between CBM-Z- and neural network-predicted concentration changes among all pollutants (MSE units: $ppb^2$), using stochastic gradient descent[51] with the Adam optimizer[52] and a learning rate that decreases by a factor of ten after each quarter of the training steps. We use a limited number of experiments to select a network size of four residual blocks (17 total fully-connected layers), an initial learning rate of 0.00128, a batch size of 128, and a training duration of five epochs based on performance on the evaluation dataset. The neural network and training algorithm are implemented using TensorFlow version 1.8;[53] all hyperparameters not described here are set to TensorFlow defaults. After selecting the hyperparameters, we train four instances of the neural network using the same hyperparameters (but random initial conditions for the trainable neural network parameters), and select the instance that performs best over a characteristic diurnal cycle as the final model.

Computer code for training the neural network emulator and the trained model are in the Supporting Information. The CBM-Z model is available by request from Zaveri and colleagues.[43]

## 2.3 Neural Network Testing

### 2.3.1 Computational speed

We compare model run times between CBM-Z and the neural network for simulating chemistry in one million independent grid cells during a simulation period of one hour. This number of grid cells approximately corresponds to one vertical layer of a CTM simulation over North America at $0.25°×0.3125°$ horizontal resolution or to a global simulation at $2°x×2.5°$ horizontal resolution with 72 vertical layers. We test each model on the available hardware for which the model could be configured without editing the source code. For CBM-Z we use a single CPU core; for the neural network we test three configurations: one CPU core, eight CPU cores, and one GPU (NVIDIA Tesla P100). Timing recorded here for both models includes the time required to copy data in and out, which may not be necessary in a production setting. CBM-Z was compiled with gfortran using the default optimization level of zero, and TensorFlow was used with the default optimization included in the downloadable binary runtime. All comparisons were performed on the same spring 2018-vintage Google Cloud Platform Compute Engine instance—hardware specifics are not known.

### 2.3.2 Single time step emulation performance

We train the neural network and select its hyperparameters based on performance in predicting changes in concentrations of pollutant species over the course of one hour. We will refer to this as "single time step" performance even though the CBM-Z model requires multiple integration steps to simulate a period of one hour. We evaluate the neural network by comparing its predictions to the corresponding CBM-Z predictions, using root-mean-squared error (RMSE) and the square of the Pearson coefficient ($R^2$).

### 2.3.3 Multiple time step emulation performance

Ultimately, we are interested in "multiple time step" performance, where changes in concentrations output by one integration step are used to adjust input concentrations for the next step as the model runs forward in time. A multiple time step performance test is more stringent than a single time step test because prediction errors during one step propagate to the next step and may compound over time; the multiple time step test is important because an approach that performs well for one time step but cannot sustain that performance over an extended simulation is of no practical use for typical air quality modeling needs.

We perform multiple time step tests by feeding the same initial conditions to CBM-Z and the neural network, independently running each model forward in time for a period of 24 hours. In each time step, the model estimates changes in concentration; those changes are added to the prior time's concentration estimate to obtain the next estimate, and the process repeats (i.e., conventional Euler integration). We exogenously adjust solar zenith angle to represent a diurnal cycle. We then compare the predictions of the model and the emulator for a series of four representative test cases demonstrative of typical atmospheric conditions: urban baseline ("UB"), urban baseline concentrations doubled ("2UB"), urban baseline concretions doubled but with a lower initial concentration of $O_3$ ("2UB – $O_3$"), and urban baseline concentrations divided by ten ("Rural"). Initial conditions for these test cases are in Table S3. In addition to the four representative cases, we perform simulations for a series of 10,000 randomly generated initial conditions produced in the same manner as the above training and evaluation data. For the randomly generated comparisons we use error—defined as the absolute value of neural network predictions minus CBM-Z predictions—to quantify neural network performance in replicating CBM-Z predictions. We pay special attention to concentrations of ozone ($O_3$), as it is a photochemically sensitive pollutant, a known human health hazard, a US EPA criteria pollutant, and in many cases is the motivation for performing an air quality simulation.

## 3. RESULTS

### 3.1 Computational performance

Figure 2 displays timing results from running the CBM-Z chemical mechanism one million independent one-hour time steps using CBM-Z running on a single CPU core, and the neural network running on one CPU core, eight CPU cores, and one GPU. For these conditions, the neural network is ~250 times faster than CBM-Z using the same hardware. The neural network running on the GPU is ~4250 times faster than CBM-Z running on one CPU. The neural network could also be run using multiple GPUs connected to a single motherboard; this would be expected to provide an additional speedup. One simulation hour is the native time step of the neural network as configured here; simulations of shorter time periods would not require less computational time. Therefore, the speed advantage of the neural network over CBM-Z decreases with decreased simulation time; for simulating five minutes, the single-core speedup factor is approximately 7 rather than 250.

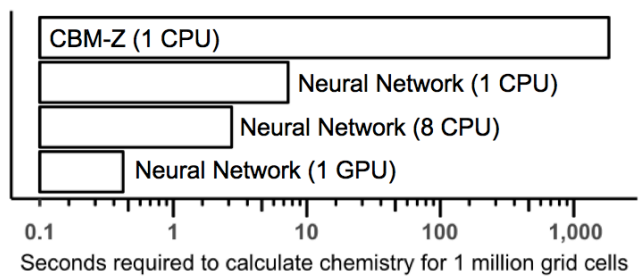

Fig 2. Time required for one million independent simulations using either CBM-Z using one CPU core, the neural network using one or eight CPU cores, and the neural network using one GPU. Model run times are in Table S4.

**3.2 Single time step emulation performance**

Figure 3 compares CBM-Z and neural network predictions for a single one-hour time step. The median neural network to CBM-Z $R^2$ among all chemical species is 0.67. The best-performing $R^2$ is for CO and $CH_3OOH$ at 0.99 and the worst performing non-zero $R^2$ is for OH at 1e-06. (The chemical species $NH_3$ and HCl do not participate in chemical reactions in CBM-Z as configured here, so their modeled concentrations do not change. Therefore, by definition, $R^2$ for these species is zero.) The median RMSE among chemical species is 0.02 ppb; the smallest non-zero RMSE is 5e-13 ppb for $O_1^D$ (excited state oxygen) and the largest RMSE is 1.97 ppb for $O_3$. All performance results are in Table S5.

The neural network is designed in a way that prioritizes prediction accuracy for chemical species with relatively large changes, so for species rates of change that are small or zero (e.g., $O_1^D$, $O_3^P$ (ground state oxygen), HCl, $NH_3$) relative measures of prediction accuracy (i.e., $R^2$) show relatively poor performance but absolute measures of accuracy (i.e., RMSE) show relatively good performance.

Figure 3 shows that for all species, the vast majority of all test cases result in small changes (represented by the black dots at the origin in each plot), while a minority of cases result in large changes (represented by the grey areas surrounding the black dots). This represents a challenge for the optimization algorithm, which works best when input and output data are normally distributed.

As a sensitivity analysis, we experimented with training the neural network against CBM-Z outputs with Gaussian normalization applied. As shown in Fig. S1, results showed substantial improvement in relative error (median $R^2$: 0.996) but degradation in absolute error (median RMSE: 0.1 ppb).

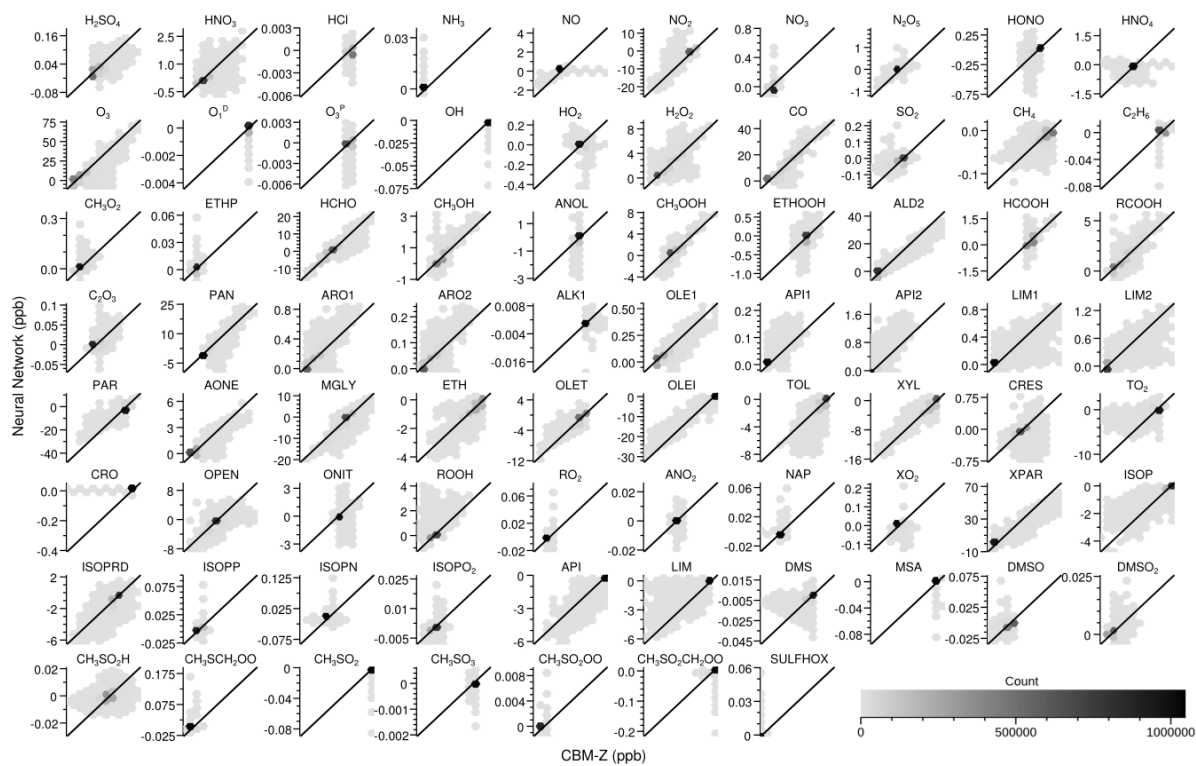

Fig. 3. Comparison of CBM-Z (x-axis) and neural network (y-axis) predictions of pollutant concentration changes after one simulation hour for 1 million example cases. The grey scale represents the number of samples in each hexagonal region; the color scheme is chosen to emphasize outliers. By design, model performance is better for species with large concentration changes than for species with small changes.

## 3.3 Multiple time step emulation performance

Figure 4 shows the evolution of the chemical system over a period of 24 hours, as predicted by

CBM-Z and the neural network, after being initialized with the urban baseline (UB) initial

condition scenario described above. The neural network steps forward in time in one-hour

intervals using Euler integration, so errors in one pollutant early in the simulation can lead to

errors in other pollutants later in the simulation. In this test case, after simulating 24 hours, the

neural network predicts $O_1^D$ concentrations with the smallest RMSE (3e-06 ppb) and $O_3$

concentrations with the largest RMSE (21 ppb). Accuracy results for all test cases can be found in Table S6.

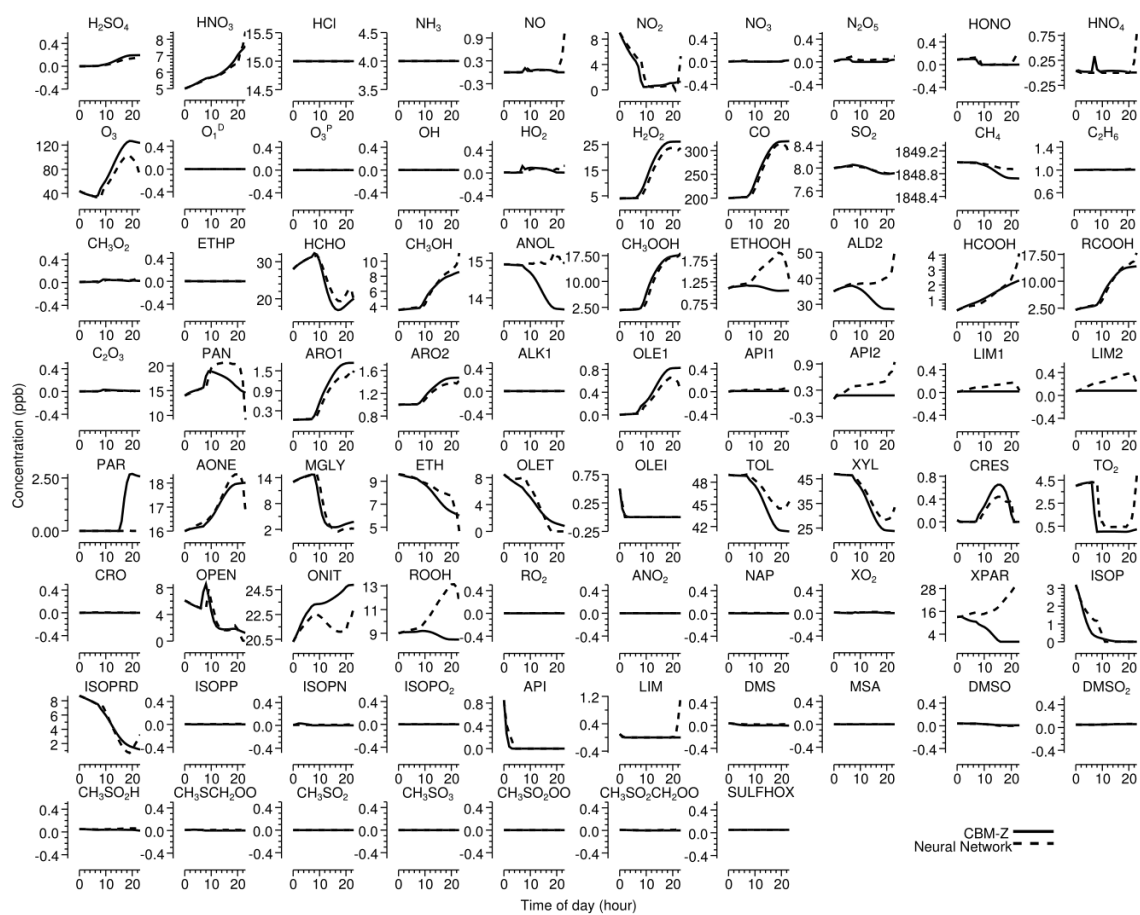

Fig. 4. CBM-Z vs. neural network comparison of simulated diurnal patterns based on representative urban baseline (UB) test case concentrations.

Figure 5 shows O$_3$ diurnal concentration results for the four test cases described in Section 2.3.3. Results for all pollutants are in Figs. S2-S4. In these cases, the neural network is generally able to reproduce the diurnal patterns (e.g., O$_3$ concentrations start increasing at ~6:00 when the sun rises and decreasing at ~20:00 when the sun sets). However, performance varies among test cases, with RMSE ranging from 21-58 ppb. As noted above, error generally increases as simulation time passes.

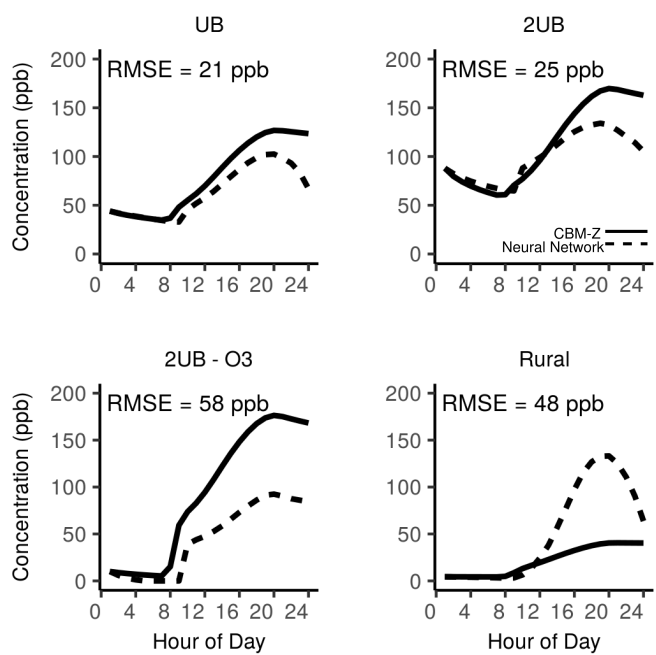

Fig. 5. Comparisons of CBM-Z and neural network simulated diurnal O$_3$ concentrations for representative initial conditions. The neural network tends to emulate general concentration trends but with increasing error over time.

Figure 6 shows the median, interquartile range, and absolute range in neural network error values for diurnal simulations of O$_3$ initialized from 10,000 randomly generated initial conditions. After 24 hours, the $0^{th}$, $5^{th}$, $25^{th}$, $50^{th}$, $75^{th}$, $95^{th}$, and $100^{th}$ error percentiles are 0.02, 7, 29, 54, 89, 190, and 280 ppb respectively. Although there are encouraging cases where the neural network matches CBM-Z almost exactly, and the median error values are low enough that fine-tuning the existing architecture and hyperparameter values may be enough to reduce them to a useful range, the existence of extremely high error values in a minority of cases is a problem that will need to be addressed before neural networks can be considered for widespread use in chemical transport modeling. Figure S5 contains corresponding plot results for all chemical species.

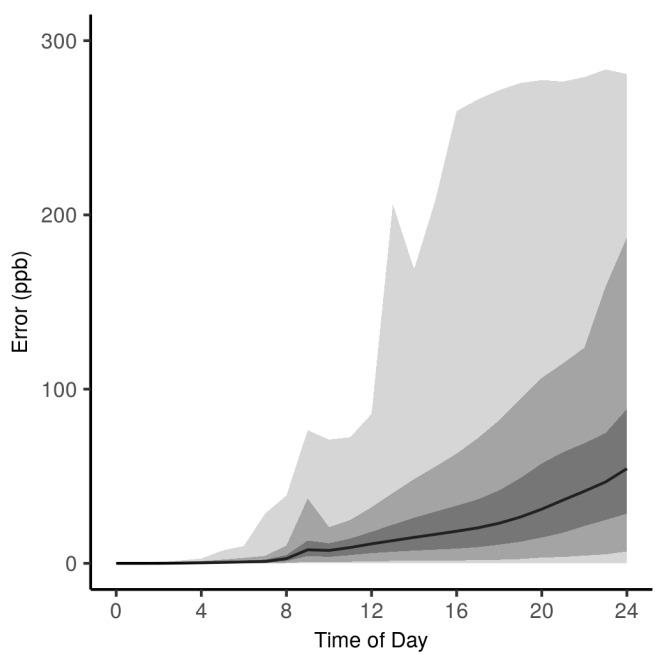

Fig. 6. Median (black line), interquartile range (darkest grey area), 5-95 percentiles (medium grey area), and absolute range (lightest grey area) of neural network vs. CBM-Z error values in $O_3$ diurnal cycles initialized with 10,000 randomly generated initial conditions. The maximum observed error must be dramatically reduced before this approach is viable for use in chemical transport models.

**4. DISCUSSION**

This work demonstrates that the use of machine learning to accelerate atmospheric chemistry computations—and potentially to model other physical phenomena—is a promising area for future research in spite of limitations in current emulative performance. We present results showing an orders-of-magnitude reduction in time required to simulate chemical reactions—the most computationally intensive component of a CTM—by emulating a large-scale (77 species) gas-phase chemical mechanism using a neural network framework. To our knowledge, this type of emulation has not been previously published. With the preliminary approach here, we are able to match CBM-Z predictions of changes in concentrations from randomly generated starting points over the course of one hour with RMSE of 1.97 ppb or less (median species RMSE of 0.02 ppb) while achieving a ~250 times computational speedup using the same hardware.

We achieve much larger speedup factors by taking advantage of modern hardware in ways that could not be easily reproduced in conventional chemical mechanisms. Although traditional chemical mechanisms such as CBM-Z are routinely parallelized to multiple CPU cores using either shared-memory (e.g., OpenMP) or distributed-memory (e.g., MPI) parallelism to speed up computation,[54] doing so can lead to bottlenecks in communication among processes and adds complexity to the software. Models of gas-phase chemistry can be rewritten to use GPU hardware,[17] but it may not be practical to do so for aerosol-phase chemistry models which typically include conditional statements and other operations that cannot be efficiently performed by a GPU. A strong advantage of neural network emulation is that it does not have this computation-flow limitation.

Additional work is necessary, however, before neural networks can be used as replacements for traditional atmospheric chemistry mechanism calculations. The main limitation identified here is that, even if in the vast majority of cases the neural network makes predictions that are relatively close to those of the target model, the neural network will occasionally make a prediction that is very different. Even with a small number of these large errors, as the simulation runs forward the errors can propagate and grow, eventually yielding meaningless predictions.

One reason for the success of neural network-based predictions in other fields has been the ability to design the network architecture to take advantage of symmetry or structure in the system being modeled, for example the fact that a cat is a cat no matter where in an image it appears[55] or the fact that a molecule can be represented as a graph of nodes connected by edges[56]. Future work may be able to resolve the exploding error issue described here by taking advantage of symmetries or structure in the atmospheric chemical system, for example the principal of conservation of mass.

**SUPPORTING INFORMATION**

Computer code for training the neural network emulator and the trained model (CSV and python scripts) and supplemental text and figures (PDF). This material is available free of charge via the Internet at http://pubs.acs.org.

**ACKNOWLEDGEMENTS**

The authors thank Peter Adams, Jiawei Zhang, Christoph Keller, and Nathan Kutz for valuable discussion, and Rahul Zaveri for use of and advice regarding CBM-Z. This article was developed under Assistance Agreement No. RD83587301 awarded by the U.S. Environmental Protection Agency. It has not been formally reviewed by EPA. The views expressed in this document are solely those of the authors and do not necessarily reflect those of the Agency. EPA does not endorse any products or commercial services mentioned in this publication.